\documentstyle[aps,prb,epsfig]{revtex}

\begin{document}
\draft
\twocolumn[\hsize\textwidth\columnwidth\hsize\csname
@twocolumnfalse\endcsname

\title{Hidden degree of freedom and critical states in a two-dimensional electron gas\\ in the presence of a random magnetic field}
\author{Hoang K. Nguyen}
\address{Department of Physics and Astronomy, University of California, Los Angeles, CA 90095--1547}
\date{\today}
\maketitle

\begin{abstract}
We establish the existence of a hidden degree of freedom and the critical states of a spinless electron system in a spatially-correlated random magnetic field with vanishing mean. Whereas the critical states are carried by the zero-field contours of the field landscape, the hidden degree of freedom is recognized as being associated with the formation of vortices in these special contours. It is argued that, as opposed to the coherent backscattering mechanism of weak localization, a new type of scattering processes in the contours controls the underlying physics of localization in the random magnetic field system. In addition, we investigate the role of vortices in governing the metal-insulator transition and propose a renormalization-group diagram for the system under study. 

\vspace{0.5cm}
\end{abstract}
]

\narrowtext

\section{Introduction}
\label{sec:introduction}
The problem of electron motion in the presence of a random magnetic field in two dimensions has been of fundamental interest in several physical situations: unitary symmetry of localization~\cite{PALee}, gauge-field description of high-$T_c$ superconductors~\cite{gauge}, and Chern-Simon theory of half-filled quantum Hall systems~\cite{anomaly}. As a result, a great amount of effort has been made during the past two decades in order to understand the transport properties of the random magnetic field (RMF) model. According to the scaling theory of localization, in the absence of interactions, all states of a two-dimensional (2D) disordered system are localized~\cite{Abrahams}. Field-theoretical study based on the nonlinear $\sigma$ model approach subsequently confirms this picture and indeed predicts that the result also holds for systems in which the time-reversal symmetry is broken~\cite{Hikami}. The RMF model, therefore, has a delicate standing, for, on the one hand, it is expected to represent the unitary class of localization in which the time-reversal symmetry is broken by a magnetic field; on the other hand, the presence of magnetic field might, arguably, give rise to some exotic effects of non-perturbative origin. Typical examples of these effects include the formation of Landau levels which cannot be obtained at any order of perturbation expansion in terms of magnetic field strength, and the celebrated topological term of the integer quantum Hall effect (QHE)~\cite{Levine}.

The fundamental and interesting question of whether 2D electrons can become delocalized in a RMF remains controversial, however. The unusual single particle properties of the model have been discussed in Refs.~\cite{Pryor,RF-DOS}. In a large number of papers, it has been argued that there is a band of delocalized states~\cite{Pryor,RF-metal,Kalmeyer,Zhang,ShengBhatt,Xie,Sheng,Taras-Semchuk}. Similarly, in a large number of papers, it has been argued that all states are localized~\cite{Sugiyama,Aronov,DKKLee,Kim,RF-nometal,Mirlin}. There are also papers in which evidence for a single critical energy has been presented~\cite{RF-critical}. Furthermore, of the papers that support the existence of conducting states, the physical origin of the delocalization in the RMF problem has not been fully addressed. Nevertheless, among these various conflicting results and conclusions, the most notable idea, to us, is the implication of a possible additional degree of freedom hidden in the RMF problem~\cite{Zhang}. Briefly stated, it was suggested that a new type of scattering specified by some hidden variable could produce or diminish a mass gap that, in turn, decides the phase of the system~\cite{Zhang}. Interestingly, an introduction of such an extra variable would, in fact, help resolve several issues related to the interpretation of the numerical results available to date.

Stimulated by this insight, {\it we have carried out a comprehensive search for a hypothesized hidden degree of freedom} that has not been encountered and explored in the context of two-dimensional localization. The purpose of our search was three-fold: to determine the existence/nonexistence of the new degree of freedom; identify its nature and origin, if it exists; and investigate its role in governing the metal-insulator transition in the RMF problem.

The magnetic field configurations considered in our work were chosen to be correlated over a finite range. There are several reasons for such a choice. Firstly, unlike the diagonal disorder case where the scalar potential fluctuation can be taken to be $\delta$-correlated, being associated with a vector potential, the magnetic field can only vary smoothly over a finite range. Secondly, it is this type of RMF that has been asserted to possess a metal-insulator transition in recent numerical work~\cite{Sheng}. Furthermore, a model with spatialy-correlated magnetic field could be connected to an effective network model; the connection, in turn, explains the nature of localization/delocalization in the RMF problem. Finally and most importantly, a smoothly varying magnetic field, physically speaking, is expected to radically influence the electron eigenstates in a similar way that a uniform magnetic field does a free electron gas to form Landau levels, opening a possibility for non-perturbation effects to come into play.

The structure of our paper is as follows. Sec.~\ref{sec:scaling} presents our numerical computation of localization length for RMF systems. In Sec.~\ref{sec:twoparameter} we propose a two-parameter scaling procedure to analyze the data. The next two sections contain our study of electron wavefunction in a RMF. The relevance of the problem under study with respect to a network model is dicussed in Sec.~\ref{sec:discussion} and followed by a summary and an appendix.

\section{Extended states in a random magnetic field and the shortcomings of the standard scaling scheme}
\label{sec:scaling}
We consider the model of non-interacting spinless electrons hopping on a square lattice subjected to a perpendicular random magnetic field and a random scalar potential. The model Hamiltonian is defined as follows:
\begin{equation}
H=-t\sum_{\langle ij\rangle}(e^{i\theta_{ij}}c^\dagger_i c_j+e^{-i\theta_{ij}}c^\dagger_jc_i)+\sum_i V_i c^\dagger_i c_i
\end{equation}
where $c_i^\dagger$ is a fermion creation operator at site $i$ and the first summation is over nearest neighbors. Here on, we shall set the hopping element $t$ to be the unit of energy and the lattice constant $a$ the unit of length. The diagonal disorder is introduced through the randomness in $V_i$ and uniformly distributed in the interval $[-\frac{1}{2}W,\frac{1}{2}W]$. The magnetic flux through each plaquette is equal to the sum of the Peierls phases $\theta$'s along its four edges. For systems with an open boundary, the Landau gauge can be chosen, namely, $\theta_{ij}=0$ on the horizontal links of the lattice. We are interested in the case where the fluxes at different plaquettes are correlated over a length scale $\sigma_f$. The flux through plaquette $p$ is then generated in the following way:
\begin{equation}
\phi_p=\frac{h_0}{\sigma_f^2/4}\sum_q f_q e^{-\frac{|R_p-R_q|^2}{\sigma_f^2}}
\label{eq:flux}
\end{equation}
where $h_0$ is to adjust the flux strength; $f_q$ is the field ``source'', chosen randomly within $[-1,1]$. Unless specified otherwise, the flux parameters in this paper are set to be $h_0=1.0$ and $\sigma_f=5.0$, corresponding to a smoothly varying flux between $-\phi_0/2$ and $\phi_0/2$, where $\phi_0$ is the flux quantum. It is easy to see that with the above set up, $\langle\phi_i\rangle=0$ and $\langle\phi_i\,\phi_j\rangle\sim \exp\left(-\frac{|R_i-R_j|^2}{2\sigma_f^2}\right)$.

In this section, we shall employ the standard transfer matrix method~\cite{MacKinnon}, which has been widely used with great success in the study of localization. The system under consideration is of strip geometry of width $M$ and length $L$, where periodic boundary condition is imposed across the strip width and $L$, in principle, has to be sent to infinity, leaving $M$ the only characteristic length for the system. In practice, $L$ is chosen to be about $10^5$ {\it longer than} $M$ to achieve the self-averaging and a first standard deviation of $0.5\%$ for data presented below. For such a large value of $L$, a successive multiplication of the transfer matrices converges and is characterized by a set of $M$ Lyapunov exponents, which determine how fast the wavefunction is damped along the strip. The smallest of these exponents, therefore, contributes the most to the transport. Its reciprocal, which has dimension of length, is the localization length $\lambda_M$ of the wavefunction, being confined on the strip. As the system width is enlarged, the localization length approaches its bulk value $\xi\equiv \lim_{M\rightarrow\infty}\lambda_M$, which has usually been named the correlation length. There are two possibilities: (i) $\xi$ is finite, in which case the wavefunction at large distances has an exponential form and the system is in the insulating phase, and (ii) $\xi$ is infinite, in which case the wavefunction is extended and the system is in the conducting phase. In order to quantify systems of finite size, one defines a {\it dimensionless} quantity, the so-called reduced localization length $\Lambda_M$, by normalizing the localization length with respect to the system width: $\Lambda_M\equiv~\lambda_M/M$. The newly defined length would serve as a suitable indicator to how the wavefunction in a disordered sample responds as the restriction on sample size is gradually relaxed, from which the meaningful information with regard to the bulk properties of the system can be extracted efficiently.

\begin{figure}[t]
\centerline{\epsfig{file=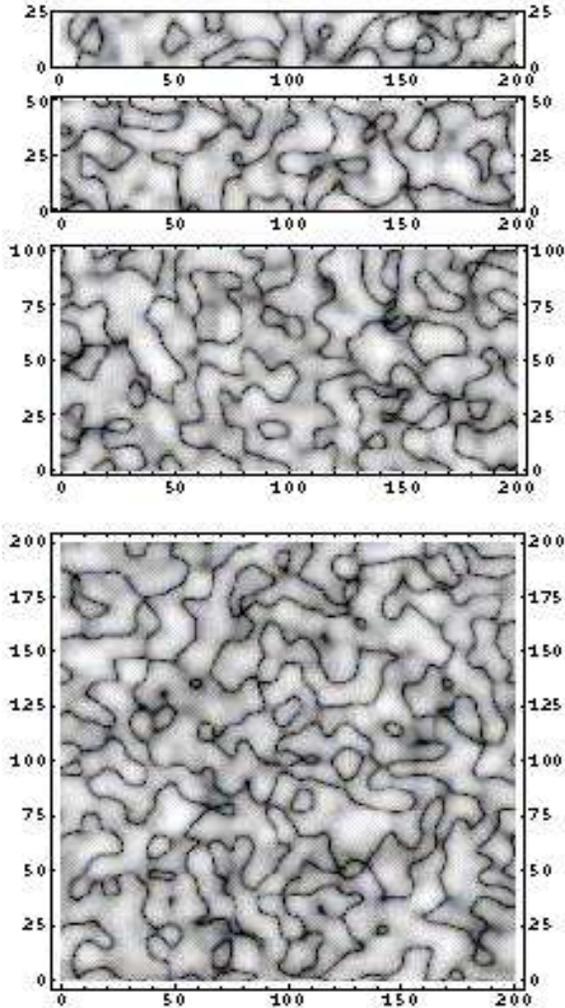,height=7.2in,width=4in,angle=0}}
\vspace{0cm}
\vskip -3cm
\caption{From top to bottom: Percolation patterns of the $B=0$ contours of the magnetic field landscape generated from Eq. (\ref{eq:flux}) for four samples of width $M=25,50,100,200$, respectively. In each sample, the top and bottom edges are connected forming a horizontal cylinder. Gray scale is based on the logarithmic of the intensity of the magnetic field. Higher fields correspond to lighter regions. Black lines represent the magnetic boundaries, which only begin to percolate in large samples.}
\label{percolation}
\end{figure}

It appears straigthforward to generate the correlated random magnetic flux as described in Eq.~(\ref{eq:flux}). Since the field configuration has zero average, fluxes of opposite sign are equally likely. However, we find it necessary at this stage to cite an advance result, which is to be shown later, that the electron transport along the strip is mainly carried by the boundaries between magnetic domains, or the zero-field contours. Thus, the implementation of flux mentioned above, although straightforward, is problematic because it violates the percolation property of these contours if the system is not sufficiently wide. In Fig.~\ref{percolation}, the magnetic boundaries are shown in black lines for different magnetic configurations produced from Eq.~(\ref{eq:flux}) with $\sigma_f=5.0$ in four cylidrical samples of successively doubled widths. Evidently, in order for the black lines in the pictures to percolate, the sample width $M$ has to reach some certain value, e.g., $100$ lattice spacings or more. Otherwise, starting from, say, the left end of the sample, an electron traveling along the zero-field contours can hardly reach the right end. Increasing the sample size will certainly cause these contours to percolate, leading to a spurious enhancement of the electron conduction. We have determined that this enhancement of the reduced localization length $\Lambda_M$ can be substantially large, in some case as large as a factor of $10$, as $M$ increases from $10$ to $240$ lattice spacings. As a matter of fact, this is what was observed in a recent numerical report~\cite{Sheng} and a false metal-insulator transition with two distinct scaling branches followed, as a result. Obviously, only when these contours become percolating thoughout the sample do the data reflect the true conduction of the system. It is thus very important to enforce the percolation condition for the zero-field lines in all the samples in use. We achieved this goal by regularly imposing a vanishing total flux through each $M\times M$ square segment in the elongated strip. In doing so, we forced domains of opposite magnetic field to cover essentially same areas, making the domain boundaries smoothly connect with one another within each segment and thus along the strip. This regulation in our flux implementation effectively removed the spuriousness inherent in the data and, as we shall see below, leads to non-trivial results.

\vskip -.75cm

\begin{figure}[htb]
\centerline{\epsfig{file=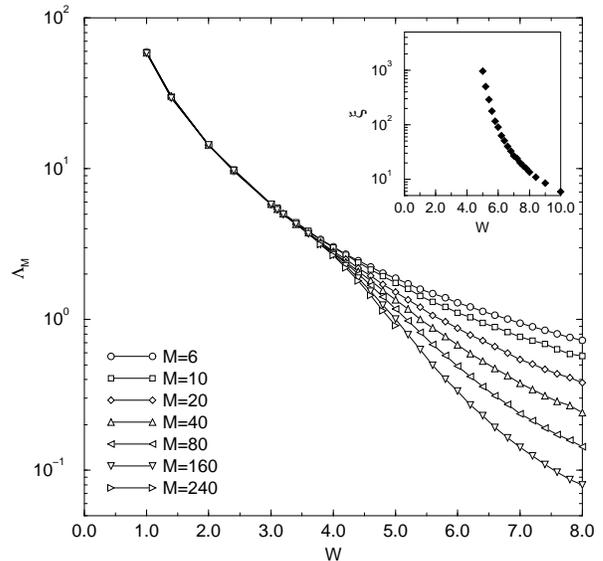,height=3.5in,width=3.5in,angle=0}}
\vspace{0cm}
\vskip 0cm
\caption{Reduced localization length in random field model with $E_F=-1.0, h_0=1.0, \sigma_f=5.0$. Inset: Correlation length obtained by data collapse (see explanation in text).}
\label{RMF_W}
\end{figure}

The main panel of Figure~\ref{RMF_W} shows our results for the reduced localization length at energy $E=-1.0$ at different values of size and disorder strength. We were able to carry out the most extensive study of this kind allowed in current computers: systems of width up to $240$ lattice constants were used at, for a reason to be clarified later, disorder of intermediate strength, $3.8\leq W\leq 5.0$. At all other disorder values, the width was chosen to be $6, 10, 20, 40, 80,$ and $160$. As can be seen from Fig.~\ref{RMF_W}, for strong disorder, the reduced localization length continuously decreases as a function of the strip width, which is to be extrapolated to zero at large $M$ limit, corresponding to an insulating phase. At very large disorder, the reduced localization length indeed experiences an exponential drop as $M$ increases, a typical behaviour in strong localization regime. Remarkably, for all small values of disorder, $W\alt 4.0$, the reduced localization length appears to be size-independent, at least within the statistical error bars due to sampling. This behavior signals the onset of a critical phase in which the electron wavefunction is self-similar at all length scales that are set by $M$. In other words, in this scale-invariant regime, the wavefunction at large distances has an algebraic spatial dependence, instead of an exponential decay. This very interesting situation has actually been encountered in several numerical studies where a line of critical points was reported~\cite{Kalmeyer,Xie}. However, the restriction in small system size and low-precision data precluded a conclusive statement to be reached in previous work. More importantly, little effort has been made to elucidate the physics of delocalization in general and of a line of fixed-points in particular for the RMF problem.

We must note that a highly plausible mechanism for delocalization in systems containing RMF can easily be seen from the very construction of our computation, in which the special implementation of flux plays a crucial role. For field configurations of vanishing mean, it is the zero-field contours that carry extended states by percolating throughout the system. When a plain application of Eq.~(\ref{eq:flux}) was used which, as is evident in Fig.~\ref{percolation}, did not ensure the percolation of these contours, the size-invariant behavior at small disorder was not observed as in Ref.~\cite{Sheng}. Clearly, by inherently imposing the percolation condition, our flux implementation has captured the essential physics of delocalization in the RMF model and has thus yielded extended states as an immediate consequence. From this explanation, the notion of a critical phase appears to be very robust and inevitable.

Perhaps rather surprising is, instead, the presence of an insulating phase in a RMF, given the percolation of the zero-field countours regardless of all other parameters in the problem. It is our main purpose in the rest of this paper to assess a legitimate, and even more fundamental, question: {\it Why localization in a RMF system?} This system deserves a thorough consideration in its own stand because the presence of a magnetic field, uniform or nonuniform, quenches the kinetic energy and radically alters the character of electron eigenstates. While coherent backscattering is the mechanism of localization in a random scalar potential, the localization in the case where the kinetic energy is quenched, like the QHE for example, is the quantum hopping of cyclotron motions. One can now envisage a situation in which electrons are subjected to a staggered magnetic field. The electrons are forced to reside mainly around the magnetic boundaries where the field changes its sign, forming what so-called ``edge states''. The introduction of randomness in the magnetic field, as well as an additional random scalar potential if applicable, scatters these states, presumably, in a fundamentally different manner from that of the Anderson localization. In other words, since the edge states are non-perturbative in essence, it is an open question of whether their character induced by magnetic field survive the introduction of disorder.

Nevertheless, in the remaining of this section, we wish to focus on the attempts to incorporate the RMF data into the general framework of the localization understanding. A common practice in the study of localization is the data collapse, in which the reduced localization length is assumed to satisfy the finite-size scaling {\it ansatz}:
\begin{equation}
\Lambda_M(W)\equiv\frac{\lambda_M(W)}{M}=f\left(\frac{\xi(W)}{M}\right)
\label{eq:ansatz}
\end{equation}
where $f$ is a universal scaling function, and $\xi=\lim_{M\rightarrow\infty}\lambda_M$ is the bulk localization length or the correlation length. As an {\it ansatz}, Eq.~(\ref{eq:ansatz}) then requires a justification which will be achieved if all the data points of $\Lambda_M$ at different values of disorder and length scale can be collapsed onto a single curve described by $f$. The correlation length is also obtained in this data collapse process and will usually be fit with some trial singular function to determine the critical point. Since the above {\it ansatz} accepts $\xi/M$ as the only argument, it is equivalent to the one-parameter scaling hypothesis that conductance is the only relevant scaling variable in the problem and that the notion of a $\beta$-function $\beta(g)\equiv d\ln g/d\ln M$ is well-defined~\cite{Abrahams}. In fact, in many localization problems, like the Anderson model in two and three dimensions, the single-parameter scaling scheme has been exploited with remarkable success~\cite{MacKinnon} and the scaling hypothesis has been verified with great confidence in these circumstances. If this scaling procedure is to be adopted and applied to our RMF data obtained in previous calculations, the correlation length for strong disorder is shown in the inset of Fig.~\ref{RMF_W}, which seems to reveal a divergence at a critical disorder of intermediate strength. However, there is a caveat. On the one hand, the scaling assumption (\ref{eq:ansatz}), by virtue of its inherent size-dependence through $\xi/M$, obviously fails to account for the scale-invariance at $W\alt 4.0$, unless $\xi$ is set to be infinite in the whole weak disorder region. On the other hand, contradictorily, data collapse cannot describe such a continuous region of infinite $\xi$ because the procedure could at most give rise to {\it isolated} fixed-point(s). The reason is that, inferred from its analyticity, the $\beta$-function could only have isolated node(s), some of which characterizes the transition point(s), like the $3D$ Anderson problem or in the presence of spin-orbit scattering.

In fact, {\it in order to successfully describe a situation with a continuum of fixed-points, it is necessary to adopt an extra scaling variable and, consequently, abandon the one-parameter scaling hypothesis}. A clear example is the Kosterlitz-Thouless transition in an $O(2)$-spin model in which the low-temperature phase corresponds to a line of critical points. Beside the spin stiffness, a new degree of freedom -- the ``fugacity'' of vortices -- is required to describe the transition. Litterally, in dealing with such two scaling variables, the examination of the $O(2)$-spin model indeed requires a special analysis~\cite{Schultka} which is briefly explained in what follows. In the ordered (low-temperature) phase, the scaling function, i.e., the scale-dependence of spin stiffness, was found by solving a set of two renormalization-group (RG) equations. At several (low) temperature values, the scaling fucntion was then fit with simulational data and extrapolated to infinite system where one could extract the bulk spin stiffness which was then used to determine the critical Kosterlitz-Thouless temperature $T_{KT}$. In the disordered (high-temperature) phase, where it was much less transparent to find a compact solution to the RG equations, the single-parameter scaling {\it ansatz} was assumed and $T_{KT}$ was calculated by data collapse procedure analogous to the localization cases. However, the critical temperature $T_{KT}$ obtained by approaching from the high-temperature side is observed to be far worse than that computed from the low-temperature side~\cite{Schultka} because the presence of the second variable necessarily invalidates the scaling {\it ansatz}.

Similarly, it should be clear that a plain application of the scaling assumption (\ref{eq:ansatz}) into the RMF case is problematic and, indeed, does not yield much reliable information with regard to the transition point. For the time being, since a microscopic theory, i.e., a set of RG equations, describing the hypothesized critical lines of the RMF problem is not ready at hand, we would like to focus, instead, on two important and intimately related issues which are of broader range of applications. They are: (i) testification of the one-parameter scaling hypothesis, and (ii) determination of the existence of an additional degree of freedom, if there is any, in a general localization problem.

Before we present our method in the next section, let us briefly discuss other available conclusions in opposition to the possibility of a line of fixed-points. Generally, such a possibility has been excluded on two following grounds:

(a) A scale-invariant phase would require an identically vanishing $\beta$-function, an unacceptable requirement given the analyticity of the $\beta$-function~\cite{Sugiyama}.

(b) A scaling function for conductance in a RMF, obtained from the data collapse using Eq.~(\ref{eq:ansatz}), was claimed~\cite{Aronov} to be reasonably close to the theoretical scaling function for the unitary class of the $\sigma$-model.

Once again, the aforementioned arguments are crucially based on an implicit assumption of the single-parameter scaling hypothesis. As for (a), the introduction of an additional scaling variable can easily lead to a critical phase without violating the analyticity of the RG equations like the famous Kosterlitz-Thouless transition. Evidently, the assumption of a one-variable $\beta$-function is no longer a meaningful concept in such a case. Neither is the one-parameter scaling function quoted in (b) if conductance is not the only scaling variable in the problem. Subsequent attempts of data collapse as well as conclusions regarding the unitary symmetry can no longer be made with any reasonable confidence. In fact, in dealing with a system that contains a critical phase, data collapse overall is a wrong practice because this phase could easily be misidentified with an insulator, as will be discussed at the end of the next section.

\section{Two-parameter analysis of the localization length}
\label{sec:twoparameter}
Since the one-parameter scaling is the central theme of localization studies, the purpose of this section is to suggest a new procedure to testify this crucial hypothesis. Our idea is quite simple. In addition to the localization length $\lambda_M$, let us consider the reciprocal of the second smallest Lyapunov exponent of the transfer matrix and denote it by $\lambda_M^{(2)}$. Being the leading and subleading contributions to the electron transport and yet being two independent physical quantities, $\lambda_M$ and $\lambda_M^{(2)}$ carry the fullest and {\it complementary} information as far as the transport properties are concerned. If the single-parameter scaling applies, i.e., the dependence of the two lengths on disorder, chemical potential, and so on can be absorbed into the correlation length $\xi$, then they can be written as functions of one dimensionless argument $\xi/M$:
\begin{eqnarray}
\lambda_M(M,W,E,\ldots)&=&M\,f\left(\frac{\xi(W,E,\ldots)}{M}\right) \nonumber\\
\lambda_M^{(2)}(M,W,E,\ldots)&=&M\,g\left(\frac{\xi(W,E,\ldots)}{M}\right)
\label{eq:2eqs}
\end{eqnarray}
where $g$ is an additional scaling function. After an elimination of the only argument available $\xi/M$, the resultant relation between the two lengths is obviously a single-valued function: $\lambda_M=f\left( g^{-1}\left(\lambda_M^{(2)}\right)\right)$. If they are now plotted agaisnt each other on a scattering histogram, then all the data points must fall onto a single curve regardless of $M$, $W$, $E$, as well as other unspecified parameters of the system. Conversely, if the data points scatter over a wide region in the histogram, the elimination after Eqs.~(\ref{eq:2eqs}) must have been invalid since the two functions $f$ and $g$ necessarily contain at least one more extra argument beside $\xi/M$. In what follows, we shall actually use a set of ($\Lambda_M,\Lambda_M^{(2)}/\Lambda_M$) instead of ($\lambda_M,\lambda_M^{(2)}$), where $\Lambda_M^{(2)}\equiv\lambda_M^{(2)}/M$, for better illustrations.

\begin{figure}[htb]
\centerline{\epsfig{file=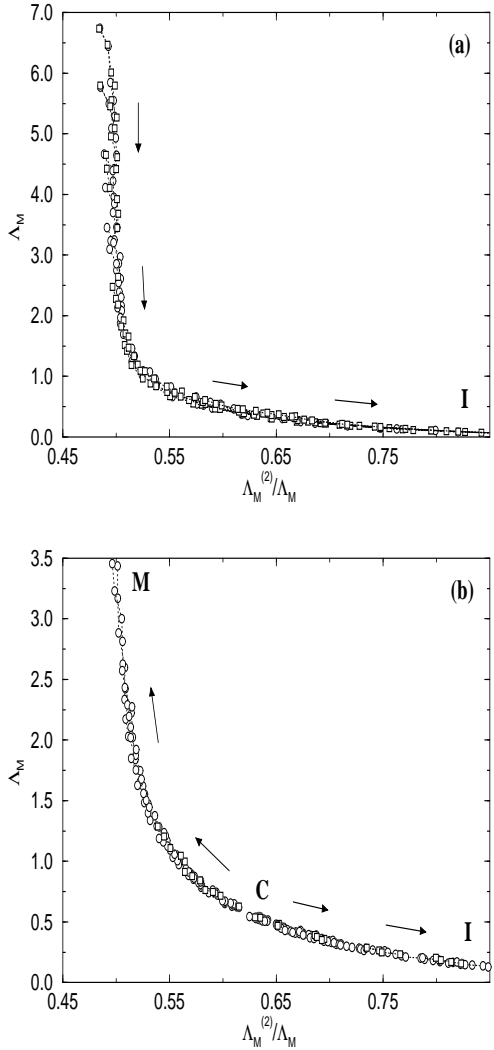,height=6in,width=3in,angle=0}}
\vspace{0.3cm}
\vskip -.5cm
\caption{Two-parameter plots of Anderson model in (a)~two and (b)~three dimensions. Fermi energy $E=0.0\ (\bigcirc)$ and $-1.0\ (\Box)$. Disorder $W$ ranges from $4.0$ to $20.0$ in~(a) and from $4.0$ to $2.0$ in~(b). Arrows show the flow direction of data along the increment of system width. $M=6,8,12,16,32,64,128,256$ for~(a) and linearly increases from $6$ to $20$ for~(b). Fixed-points at $\Lambda_M=0$ and at $\Lambda_M\rightarrow\infty$ are denoted by $I$ (insulator) and $M$ (metallic), respectively. $C$ indicates the isolated repulsive critical point.}
\label{A}
\end{figure}

We first apply the procedure discussed above into the well-understood $2D$ Anderson model: particles hopping from site to site on a square lattice and scattered by a random scalar potential. Theoretically, the single-parameter scaling hypothesis is believed to hold and all states are expected to be localized~\cite{Abrahams}. Figure~\ref{A}(a) shows our results of the localization length $\Lambda_M$ and the ratio $\Lambda_M^{(2)}/\Lambda_M$ for systems of different sizes and at various values of Fermi energy and diagonal disorder. Specifically, the Fermi energy is $E=0.0$ (circle symbols) and $E=-1.0$ (square symbols), and the disorder runs from $2.0$ up to $20.0$, both in the energy unit of the hopping element. Arrows indicate the direction of data as $M$ is successively increased. Remarkably, an excellent verification of the single-parameter scaling theory in the $2D$ Anderson model is obtained for systems of size as small as $6$ lattice spacings. As evident in Fig.~\ref{A}(a), all the data points {\it automatically} align on a single curve which also appears to be independent of Fermi energy. Moreover, as the system is enlarged, the data points are driven along the curve towards $I$, the insulating ``fixed point'' corresponding to $\Lambda_M=0$. That there is no other fixed point that separates the flow line is a definite indication that the system entirely lies in an insulating phase.

The method is next applied to another well-understood situation, the Anderson model in three dimensions, in which the scaling hypothesis is again believed to hold and, interestingly, a metal-insulator transition has been predicted~\cite{Abrahams}. The results are shown in Figure~\ref{A}(b) for systems of size ranging from $6$ to $20$ and at different energies and disorders. Again, with the self-alignment of data points, the scaling hypothesis is indisputably verified in this case. Interestingly, there are three fixed-points on the flow line: the insulating fixed point $I$ with $\Lambda_M=0$ (strong disorder regime), the conducting (metallic) fixed point $M$ with $\Lambda_M=\infty$ (weak disorder regime), and a {\it single} repulsive (critical) fixed point $C$ that separates the two phases. Moreover, this critical point appears to be insensitive to the Fermi energy, confirming the universality of the critical conductance in the $3D$ Anderson model.

\begin{figure}[htb]
\centerline{\epsfig{file=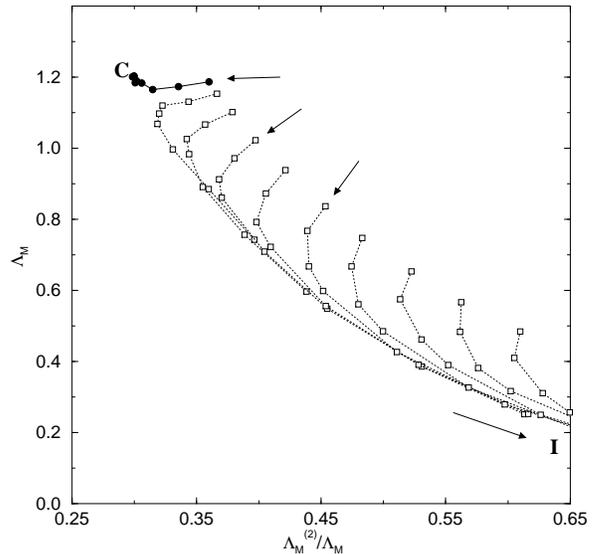,height=3.5in,width=3.5in,angle=0}}
\vspace{0.3cm}
\vskip -.6cm
\caption{Two-parameter plot of Chalker-Coddington network model, describing the quantum Hall effect at criticality $\theta_c\approx 0.881374$ ($\bullet$), and away ($\Box$) where $\theta$ ranges from $0.8$ down to $0.1$. Arrows show the flow direction of data along the increment of system size $M=6,8,12,16,32,64,128,256$. $I$ indicates the insulating fixed-points, while $C$ is the critical point.}
\label{NM}
\end{figure}

We next turn our attention to an also well-studied case: the integer quantum Hall effect. The model in use is the Chalker-Coddington network~\cite{Chalker} which is characterized by a node parameter describing the scattering between cyclotron motions across the saddle point of the potential landscape. At the center of each Landau band, where the scattering is symmetric, the node parameter is determined to be the critical value $\theta_c\equiv\ln(1+\sqrt 2)\approx 0.881374$. It has been established that the network model only sustains extended states at criticality and localized states otherwise~\cite{Chalker}. Once again, our procedure beautifully reveals in Figure~\ref{NM} the expected feature of this model: Only at criticality, $\theta=\theta_c$, the data flow runs towards and terminates at the critical point denoted by $C$; while all other flows follow a common curve to end at $I$, the insulating fixed point at $\Lambda_M=0$. The critical point $C$ governs the flow lines within its vicinity. Evidently, unlike the Anderson models presented in the previous paragraphs, the data flow of the network model do not fall on a single curve, thereby signaling the presence of an extra scaling variable. Theoretically, the quantum Hall system is controlled by another relevant variable~\cite{Levine}, the Hall conductance $\sigma_{xy}$, beside the logitudinal conductance $\sigma_{xx}$. What is striking in our picture is the evidence of a new scaling variable which, to our understanding, is directly observed for the first time without {\it a priori} knowledge of $\sigma_{xy}$. Overall, the usefulness of the method we have introduced so far lies in its capability of detecting a new scaling variable when there is one, while giving a null answer when there is none.

\begin{figure}[htb]
\centerline{\epsfig{file=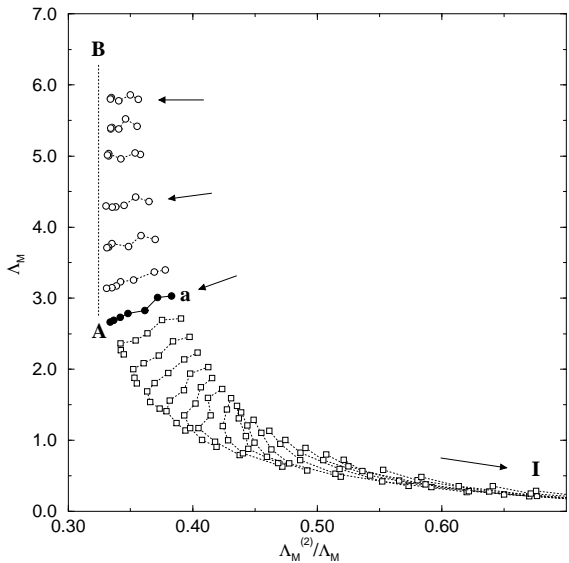,height=3.5in,width=3.5in,angle=0}}
\vskip 0cm
\caption{Two-parameter plot of the random magnetic field model at energy $E=-1.0$, field parameters $h_0=1.0$, $\sigma_f=5.0$. Disorder $W=3.0,3.1,3.2,3.4,3.6,3.8$ ($\circ$), $4.0$ ($\bullet$), and from $4.2$ to $14.0$ ($\Box$). Error bar is of size of symbols. Arrows indicate the flow direction as system size is increased $M=6,10,20,40,80,160,240$. $I$ is the insulating fixed point at $\Lambda_M=0$. Dotted line $AB$ is the locus of putative critical points. $aA$ is the separatrix between the critical and insulating phases.}
\label{RMF}
\end{figure}

Interestingly enough, the application of our analysis into the RMF model provides much appealing information. We present in Fig.~\ref{RMF} data for Fermi energy $E=-1.0$, disorder $W$ from $3.0$ to $4.0$ (circles) and from $4.2$ to $14.0$ (squares), and system width $M$ running from $6$ to $240$. Clearly, since the data points do not fall on a single curve, this must be taken as a robust evidence of a hidden scaling variable and hence a breakdown of the one-parameter scaling hypothesis in the RMF problem. As we have mentioned, it is the use of $\Lambda_M^{(2)}$ in our analysis that maps out this extra variable unambiguously, which is not allowed to observe within the standard scaling procedure. Amid its anticipatory existence that we discussed at length in the previous section, this is really the first time the new degree of freedom has manifested itself. Its pronounced influence on the RMF data, indeed, can be seen in Fig.~\ref{RMF}. There are two distinct regions corresponding to different behaviors of data, separated by the solid line $aA$ of disorder $W=4.0$. Below this line, data lie in the regime of strong disorder and are attracted towards $I$, the insulating fixed point at the lower right corner of the histogram. In contrast, data above $aA$ tend to flow towards the left side of the histogram and stop on the dotted line $AB$, a coalescence of fixed points. Clearly, this two-parameter picture has provided us a global view in determining the ultimate fate of the RMF system at large length scale. As a result, one only has to focus in the vicinity of $A$, the end point of the critical regime, to distinguish the behavior of the two corresponding phases. By the time the system size reaches $240$ lattice constants, while the line $aA$ itself and the one above it have come to termination, the line right below it, after spending sometime in this vicinity, has already meandered to the insulating fixed point. We therefore are on a solid ground to conclude that our RMF system undergoes a metal-insulator transition at $W_c=4.0$. In addition, the conductance at this critical disorder and at the largest length scale (i.e., at $A$) is determined to be $1.26\,e^2/h$, of the order of the quantum conductance, in justification for the genuine quantum-mechanical nature of this transition.

\vskip -1.5cm

\begin{figure}[htb]
\centerline{\epsfig{file=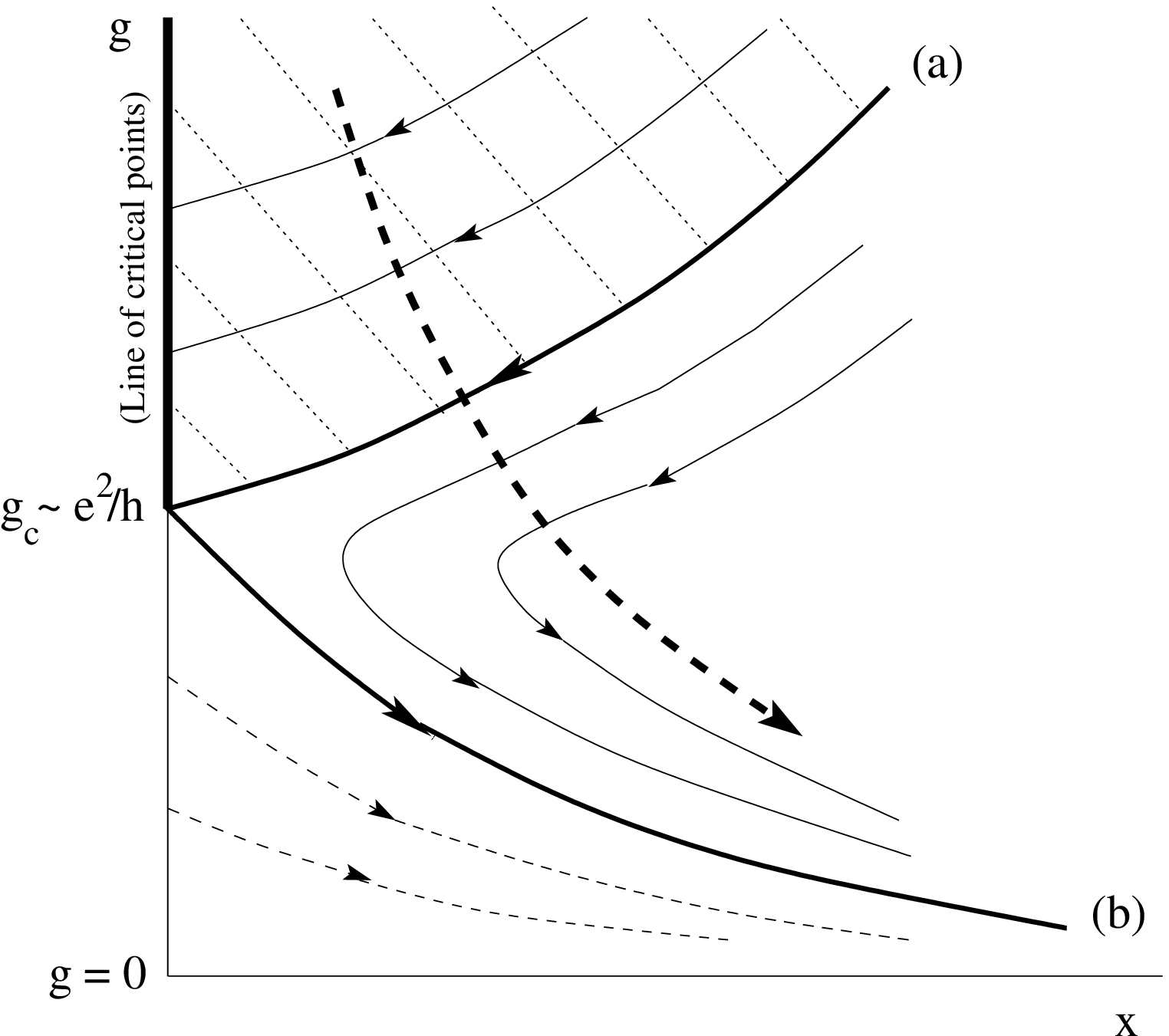,height=2.75in,width=2.8in,angle=0}}
\vskip 1cm
\caption{Schematic renormalization-group flow diagram of the random field model. $X$ represents the hidden degree of freedom. $g$ is conductance and its critical value is $g_c$. Shaded region above line $(a)$ is the critical phase. Dotted lines below line $(b)$ are projected. Arrowed thick dotted line is the projected sweep at microscopic length scale.}
\label{RG}
\end{figure}

Also note that the two-parameter pictures in all four considered situations share a common feature: their data flows do not cross. We are thus able to make a general statement. The evolution of a system with respect to an increase in size only depends upon its current state. In other words, if one attempts to construct a set of renormalization-group (RG) equation(s) for the system, its analyticity will be, in effect, ensured. For the RMF model, in account for the evidence of the hidden degree of freedom and basing on our numerical data, we propose a RG flow diagram shown in Fig.~\ref{RG}. The picture is in close resemblance to the RG diagram of the $O(2)$-spin model in which the conductance is replaced by the temperature inverse. The thick dotted line is the projected sweep at the microscopic length scale. The shaded region above line $(a)$ is the critical phase with infinite correlation length. As one moves along the RG line in this region, one always stays at criticality and the wavefunction thereby looks self-similar at all length scale; or in other words, the wavefunction has a power-law decay character at large distance. The new degree of freedom - the ``fugacity'' - is eventually renormalized to zero in this phase. Out of this region, the RG lines are driven along line $(b)$ towards the insulating fixed point where the ``fugacity'' is further enhanced.
\vskip .5cm

As our final point in this section, it is very interesting to observe that in contrast to general assumption, the reduced localization length in the critical phase is also subjected to a {\it slight size-dependence}. As evident from Fig.~\ref{RMF}, on the boundary of the critical phase (line $aA$), $\Lambda_M$ endures a $12\%$ decrease from its initial value as the system size sweeps from $6$ to $240$ lattice spacings. In account for this size-dependence, the wavefunction must acquire an additional logarithmic spatial correction, apart from the power-law form. Again, this situation has actually been well-known in the Kosterlitz-Thouless transition in which the spin-spin correlation at $T_{KT}$ is established to be~\cite{Chaikin}
\[\langle\vec S({\bf x})\,\vec S(0)\rangle\sim\frac{\ln^{1/8}|{\bf x}|}{|{\bf x}|^{1/4}} \]
In addition, the spin stiffness in its ordered phase is also modified by a logarithmic correction~\cite{Schultka}. To our knowledge, this is the first time a similar situation has been encountered in the context of localization. The implication of this slight scale-dependence is far-reaching in both theoretical and experimental sides. It, in fact, resolves a central issue in interpreting numerical data regarding the critical phase of the RMF model. Using the Landauer formula~\cite{Fisher}, the conductance at the critical disorder $W_c=4.0$ is found to be: $g(W_c,M=6)\approx~1.58\,e^2/h$ and $g(W_c,M=240)\approx~1.26\,e^2/h$. That means the conduction has undergone a considerable reduction, which is about $20-25\%$, as a function of the system size. We are confident that several numerical studies in the RMF model have mistakenly interpreted the decrease in conductance as a signature of an insulating state. In addition, due to this size-dependence, any attempt to collapse data points onto a scaling curve is no longer sensible. A similar conclusion can apply equally well to experimental interpretation where the inverse of temperature plays the role of the length scale. A drop of conductance as temperature is reduced might not necessarily imply an insulating groundstate, especially when the groundstate is critical in essence. We suggest that, if possible in such a case, another measurable quantity should be used together with the conductance to map out a two-parameter picture and identify the critical phase with greater confidence.

\newpage

\section{Anatomy of the wavefunction in a random field: qualitative pictures}
\label{sec:wavefunction}
Our examination so far has evinced the evidence of a critical phase, by virtue of the percolation picture, and a hidden hidden degree of freedom, inferred from the two-parameter analysis. Nevertheless, there is as yet no microscopic understanding of these observations. The questions to answer include: What is the nature and origin of newly found degree of freedom? What role does it play in governing the metal-insulator transition? What is the mechanism of localization in a RMF, after all? We devote this section to addressing these important issues by approaching the problem at its bottommost: the structure of its wavefunction.

We exploited the Lanczos diagonalization algorithm to compute the wavefunction of electrons on a square $300\times 300$ sample with periodic boundary condition (PBC) imposed in both directions. The sample was subjected to a smoothly varying magnetic field with zero total flux. No diagonal disorder applied, otherwise. We adopted the choice of gauge suggested in Ref.~\cite{Kalmeyer}. For the purpose of illustration, the correlation length of the magnetic flux was chosen to be $\sigma_f=15$ lattice constants, and $h_0=4.0$, correponding to a strong field. The wavefunction closest to a given energy $E$ was obtained from the Lanczos diagonalization of the $(\hat H-E)^{-1}$ matrix~\cite{Yoshino}, that is to say, the eigenstate corresponding to its largest eigenvalue in modulus was selected.

\begin{figure}[htb]
\centerline{\epsfig{file=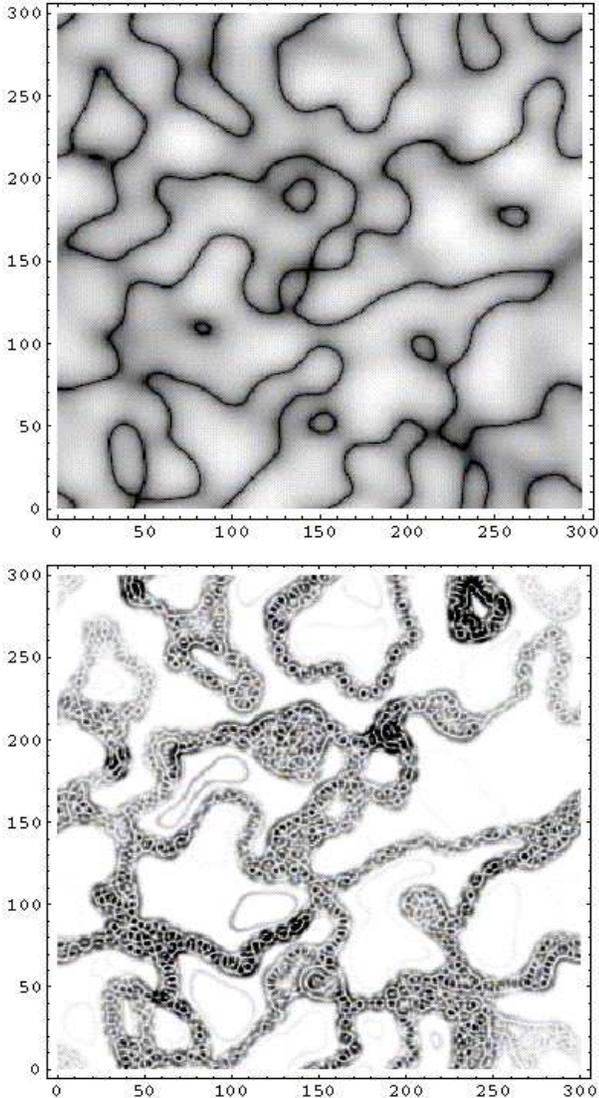,height=5.5in,width=3in,angle=0}}
\vspace{0.3cm}
\vskip 1.5cm
\caption{Upper panel: Zero-field lines of a random field (described in text). Lower pannel: Density plot of the square of a wavefunction at energy $E=-3.0$ for the above field configuration.}
\label{WF}
\end{figure}

Figure~\ref{WF} presents our results at Fermi energy $E=-3.0$, close to the band edge. The upper panel displays a typical flux configuration, in which the magnetic boundaries shown in black lines are percolating. (Note the PBC on all four sides of the sample). The lower panel shows the probability density, i.e., the wavefunction squared, where the dark spots represent a high density region. In comparing the two panels, our initial impression is that the electron wavefunction is clearly extended throughout the sample, following the fractal pattern of the zero-field contours of the magnetic field landscape. Electrons favorably reside around these lines within a few lattice spacings, forming a quasi-1D tube with finite length. We have checked that stronger field and/or higher Fermi energy would broaden the width of the tubes, but the overall picture remains unchanged. At the first sight, it is therefore sensible to describe the RMF system in the language of a one-dimensional network, similar to that proposed by Chalker and Coddington~\cite{Chalker} to represent the integer QHE. Given the inherent connection of the tubes at the saddle points of the field landscape, the scattering at the nodes of the network is symmetric~\cite{Chalker}, so the criticality of the system appears inevitable. Thus, the network picture justifies our assertion in Section~\ref{sec:scaling} regarding the relevant mechanism of delocalization in a RMF: it is the percolation of zero-field contours that is responsible for the extensiveness of the wavefunction and, hence, the existence of the conducting states.

As an interesting consequent of this picture, when an external magnetic field is applied atop the RMF, evidently, the percolation of magnetic boundaries is lifted. The RMF system is then expected to behave somewhat closer to an insulator, causing a positive magnetoresistance. This expectation is borned out by the results obtained by Kalmeyer and co-workers in a numerical work~\cite{Kalmeyer} which establishes that the RMF magnetoresistance is positive in both conducting and insulating phases. The result for the insulating phase is quite distinctive in comparison to the Anderson localization in which case magnetic field, by suppressing coherent backscattering, results in a negative magnetoresistance instead. These authors also related the magnetoresistance of the RMF model to a striking behavior of longitudinal conductance $\rho_{xx}$ of the half-filled quantum Hall system. In particular, it has been observed in experiments~\cite{Jiang} that the quantum Hall system develops a deep minimum in $\rho_{xx}$ as magnetic field is swept through its half-filling value. Theoretically, at such a strong magnetic field where the number of flux quanta is twice the number of electrons, each electron captures two flux quanta to form a fermionic electron-flux tube composite, which feels no net magnetic field. It is argued, however, that the composites are still subjected to a static random flux distribution induced from the presence of a random potential. That is to say, the RMF model is supposed to be the correct description of the quantum Hall system close to half-filling. Slightly away from half-filling, the composite fermions now experience a small magnetic field remainder after the flux quanta have been absorbed. In view of this picture, the deep minimum of $\rho_{xx}$ can be explained at once as a direct consequent of the percolation status of the zero-field contours. What we find appealing, by turning the above statement upside-down, is that the very observation of the minimum of $\rho_{xx}$ in experiments provides us an independent and complementary evidence for the extendsiveness of the wavefunction along the lines of zero field. While we are cautioned that a positive magnetoresistance is also shared by a Fermi gas, we consider this scenario unlikely as a Fermi gas would eventually become localized with an infinitesimal amount of impurities, thereby unable to account for the apparent conducting behaviour of half-filled quantum Hall systems.

\begin{figure}[htb]
\centerline{\epsfig{file=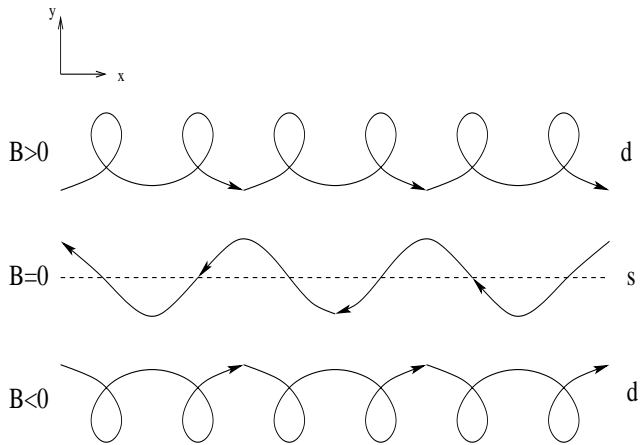,height=2.3in,width=3.3in,angle=0}}
\vspace{0.3cm}
\vskip 0.5cm
\caption{Schematic illustration of classical trajectories of drift states (d) and snake states (s) in a nonuniform magnetic field, which flow in opposite directions, in most cases.}
\label{Muller}
\end{figure}

The physical role of the $B=0$ lines mentioned above has indeed been clarified in detail by M\"uller~\cite{Muller}, which studied the electron motion in a linearly varying magnetic field. We shall reproduce its main results in the Appendix, while giving a brief explanation of the ingredients necessary for our coming discussions. At the classical level, there are two types of trajectories sketched in Fig.~\ref{Muller}. In the region far away from the $B=0$ line, the particle follows a cyclotron orbit whose guiding center drifts perpendicular to the field gradient along $\hat{\bf x}$-(forward) direction. Following M\"uller, we shall call this orbit a ``drift state''. In the region of small field $B\simeq 0$, the particle travels in a snakelike trajectory back and forth across the $B=0$ line along, most of the time, $-\hat{\bf x}$-(backward) direction. This trajectory has been named a ``snake state''~\cite{Chklovskii}, apparently in deference to its snakelike shape. Sometimes, all these trajectories are commonly called ``edge states'' for their relative position to the zero-field line. As pointed out by M\"uller~\cite{Muller}, the finite transverse size of these states is actually a purely quantum-mechanical effects since a classical orbit has no constraint on its radius. In addition, the quantum tunneling between a pair of drift (snake) states results in a new pair of symmetric/antisymmetric states, or equivalently, states of opposite parity pairwise. Detailed quantum mechanical treatments~\cite{Muller} further confirm that, in most cases, in each pair of energy bands there are two snake and two drift states at the Fermi energy. Most prominent are the snake states which peak right on the $B=0$ line; so in the context of the network model, they have a considerable chance to percolate and thus carry extended wavefunctions. The drift states, on the other hand, due to their position off the zero-field line, at the first sight, are less likely to contribute to the RMF transport. We shall soon clarify their fundamental role, however. In the meantime, we must note that there have been studies that attempted a different reasoning: the couple of snake states would ultimately become localized due to the mixing (tunneling) between themselves~\cite{DKKLee,Kim}. Firstly, the RMF wavefunction demonstrated in Fig.~\ref{WF} disagrees with this conclusion. Our careful examinations on the localization length and magnetoresistance, furthermore, depict a delocalization picture instead. More importantly, we shall point out in the Discussion that the version of the network model considered in those works was too restrictive to account for the RMF problem, unless under two major revisions. Specifically, that version assumed an incorrect scattering matrix at the nodes of the network, or the regions where the zero-field lines meet.

\begin{figure}[htb]
\centerline{\epsfig{file=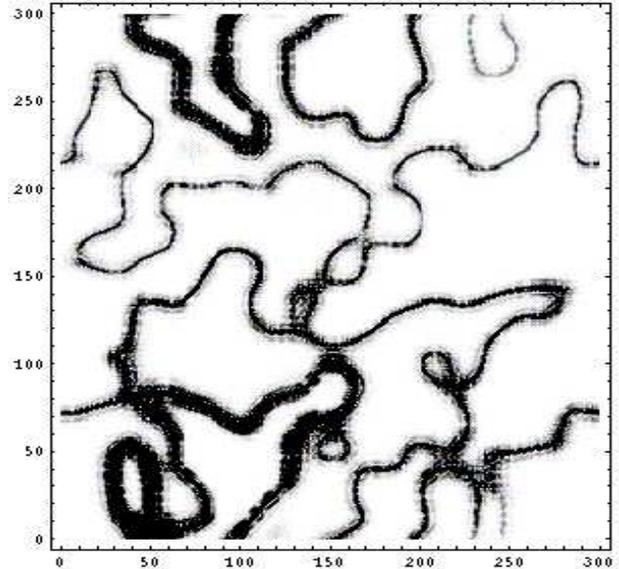,height=4in,width=4in,angle=0}}
\vspace{0.3cm}
\vskip -1.5cm
\caption{Density plot of the square of a wavefunction in the quantum Hall system. Fermi energy $E=2.8$ close to the center of the lowest Landau band and $1/4$ flux quantum is threaded through each plaquette. The diagonal disorder configuration is chosen as in Fig.~\ref{WF}(a) with $h_0$ replaced by $V_0=2.0$.}
\label{WFHall}
\end{figure}

In view of the percolation pattern, how, then, could an electron gas in a RMF become localized under some circumstances, say, at strong disorder or close to the band edge? To answer this question, it is necessary to reconsider the case of QHE where the percolation also plays a key role. As well-known~\cite{Prange}, in this situation electrons at energy $E$ reside on the equipotential lines $V=E-(n+\frac{1}{2})\,\hbar\,\omega_c$. We produce in Fig.~\ref{WFHall} a wavefunction of a potential congiguration similar to that in the upper panel of Fig.~\ref{WF}(a), with $h_0$ replaced by $V_0=2.0$. The magnetic flux per plaquette was $\frac{1}{4}\,\phi_0$, and the Fermi energy $E=2.8$, right at the center of the lowest Landau level. Again we obtain a similar picture that supports the percolation of an extended state. In this sense, the quantum Hall and RMF systems share the common physics of delocalization. However, whereas the development of insulating states in the former is achieved by destroying the percolation as one moves away from the Landau band center, this picture offers little clue towards the localization in a RMF.

Closer inspections of Fig.~\ref{WF}(b) versus Fig.~\ref{WFHall} reveal a striking difference between the corresponding wavefunctions of the two cases, however. For the quantum Hall system, aside from the percolation pattern, there is no other noticeable feature in view: the wavefunction is characterless; it is everywhere monotonous. In startling contrast, the RMF wavefunction posseses a far more complex structure along its main frame. As is prominent in Fig.~\ref{WF}(b), every segment in its network is solidly filled with series of pearl-like objects, tiny spots of high and low density succesively arranged next to one another. Apparently, this unusual feature implies some form of singularity, which, at our first guess, is the existence of zeros in the wavefunction. As a matter of fact, a definite answer to our initial speculation can be found from the examination of electron motion outlined in a previous paragraph. Let us consider again the edge states in Fig.~\ref{Muller}, as each travels along the $B=0$ line. Two noticeable aspects can be seen immediately. The snake and drift states are: (i) vastly distinct in wavenumber by flowing in opposite directions; (ii) widely far apart while maintaining a considerable overlap. It is these important differences that profoundly result in zeros (or nodes) of the wavefunction when the two states scatter against each other. Furthermore, it is an established fact~\cite{Dirac} that an isolated zero is necessarily the center of a vortex current. As an illustration, we shall give in the Appendix an explicit construction of a wavefunction with arrays of nodes and current circulating around them.

\begin{figure}[bht]
\centerline{\epsfig{file=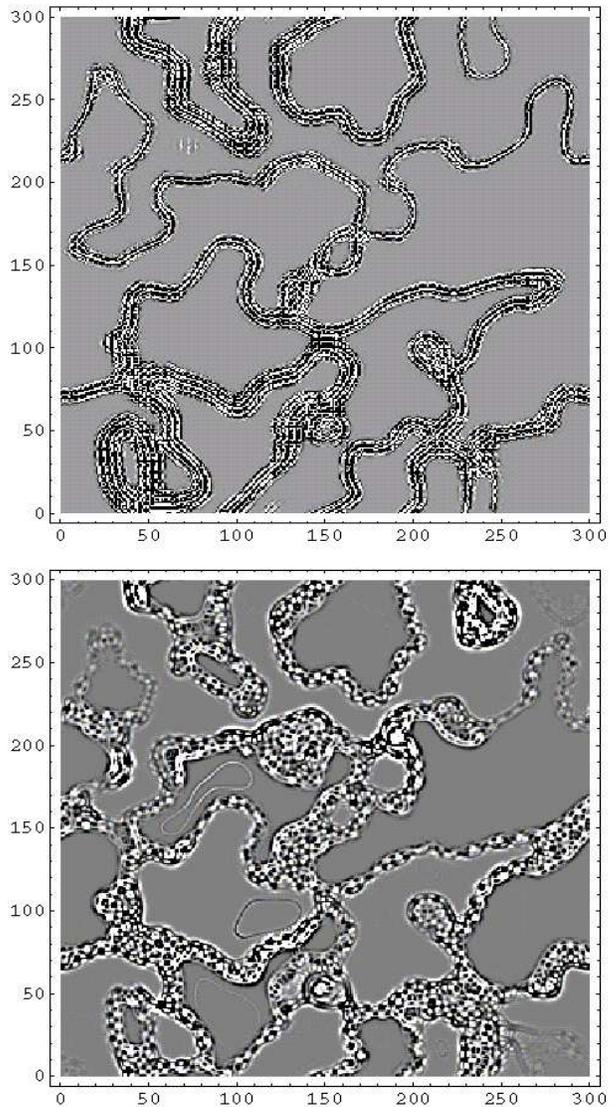,height=5.5in,width=3in,angle=0}}
\vskip 1.5cm
\caption{Curl of current, as described in text. Upper panel: quantum Hall system, using wavefunction in Fig.~\ref{WFHall}. Lower panel: RMF system, same wavefunction as in Fig.~\ref{WF}(b).}
\label{curlJ}
\end{figure}

For the time being, to visualize the current vortices in our Hall and RMF cases, we compute the curl of current flowing inside each plaquette: $(\vec\nabla\times\vec J)_\perp=\sum_\Box J_{ij}$, where the summation is around the plaquette and $J_{ij}$'s are current on its four edges. Shown in Fig.~\ref{curlJ} are the curl results for the wavefunctions in Figs.~\ref{WFHall} and~\ref{WF}(b) respectively. Once again, a salient contrast between the two cases is manifest. In the upper panel, currents within the links of the Hall network are continuous, the flow is laminar. In fact, the simplicity and regularity of current flows explain the spectacular success of the network model in representing the QHE, most notably its critical exponent. In other words, the Chalker-Coddington model has captured the relevant physics, indeed the most important one, of the integer QHE: the scattering of currents at the saddle regions of the potential landscape. In the lower panel, the RMF current pattern demonstrates a discontinous flow, the current is irregularly disrupted into arrays of vortices and antivortices represented by bright and dark spots of size about a few lattice spacings. While evident everywhere, one can easily find, e.g, on the segment at the lower left corner a series of such spots arranged in a consecutive fashion.

It appears very likely that the newly found vortices are intimately associated with the compelling evidence of the hidden degree of freedom established in the our scaling study. In fact, both phenomema should be viewed as two sides of the same coin. On the one hand, the formation of vortices is the direct consequence of the mixing of drift and snake states, which, on the other hand, the hidden degree of freedom characterizes. Accordingly, we are in a position to make the following conjecture. While coherent backscattering, i.e., the interference between an electron close path and its time reverse, is the underlying physics of weak localization~\cite{Langer}, {\it the interference between snake and drift states is the driving force of localization in systems subjected to a random magnetic field.} As a result, the RMF system can only achieve its insulating phase, under some circumstances, through the coupling processes on the links of its network, as opposed to the node scattering in the QHE situation. We must emphasize that our conclusion is not the same as those discussed in other previous work in two essential aspects. First, an electron gas in a RMF still manages to maintain its criticality, or its extended phase, if the scattering of the edge states is not sufficiently strong. There is an inherent competition between two tendencies: the delocalization, inferred from the percolation of the snake states, and the localization, induced from the interference of the snake {\it and} drift states. Second, early studies always neglected the profound role of the drift states, which is important to be restated as follows. It is only due to the vast differences in wavenumber and spatial location between a pair of snake and drift states that vortices can be pronouncedly formed.
\vskip .5cm

A successful microscopic theory for the RMF must therefore involve a correct description of process occuring on the links of the network. This observation has, in fact, been mentioned by Zhang and Arovas~\cite{Zhang}. Regardless of the technical correctness of their formulation, we are certain that various intuitive aspects of the picture drawn in their work are corroborated in our present study. Specifically, the elementary excitations of an electron gas in a RMF are found to be the edge states; states that are produced by the peculiar interplay of quantum mechanics and magnetic field. Scattering among these states gives rise to vortices, a new kind of secondary excitations. Characterizing the vortices is their ``fugacity'', the hidden degree of freedom independently observed in our scaling analysis. Depending upon physical conditions, the vortices can unrestrictedly proliferate. In the phase where the scattering is irrelevant, or few vortices are present in each link of the network, the ``fugacity'' is suppresed, the edge states remain gapless and carry currents. In the phase where scattering is relevant, or there are too many vortices involved, the ``fugacity'' is enhanced, a mass gap opens up driving the system into localized phase.

Overall, we conclude that the correct RMF network model must take into account the effect of vortices. Every link in the network is actually a quasi one-dimensional tube with vortices residing inside. Intuitively, one would expect the presence of vortices to `block' the current flows, hence driving the conduction on the contours, and ultimately, the metal-insulator transition (MIT). As a matter of fact, the possibility of such a scenario has been studied in great details in the context of quantum wires and tubes. It has been established that the formation of vortices critically affects the current flow in two-dimensional bent tubes~\cite{Wu}.

\section{Anatomy of the wavefunction in a random field: quantitative results}
\label{sec:current}
Having speculated about the possible connection between vortices and the MIT in a RMF, our task at hand is to seek out a characteristic measure to probe such a connection. Clearly, a suitable candidate is the average transmission coefficient through the links of the network. Also, what makes it attractive is that it is a local property of the wavefunction, rather than a global one like the localization length or conductance.

Let us again consider an $M\times M$ square sample wrapped on a torus and subjected to a RMF and a random scalar potential. For a finite sample, there is a non-dissipative current flowing around the torus in both of its directions. [Note that there is no edge currnt in these samples.] The value of this total current varies from sample to sample and vanishes on average since RMF configurations come in pairs that support current of opposite signs. However, its root-mean-squared (rms) fluctuations is still a sensible quantity to characterize the conduction. Also in the $M\rightarrow\infty$ limit, with no net magnetic field, the total current is of course vanishing. Therefore we expect the finite-size current to scale as $J_M\sim M^{-\nu}$, where $\nu$ is a positive number. Our naive guess is that $\nu=2$ because the total current reflects the overall conduction in all the set of links and would suffer if any of them is blocked by vortices and the number of links is proportional to $M^2$. In other words, we may interpret $M^2\sqrt{\langle J_M^2\rangle}$ as the average local current in an $M\times M$ sample.

\begin{figure}[htb]
\centerline{\epsfig{file=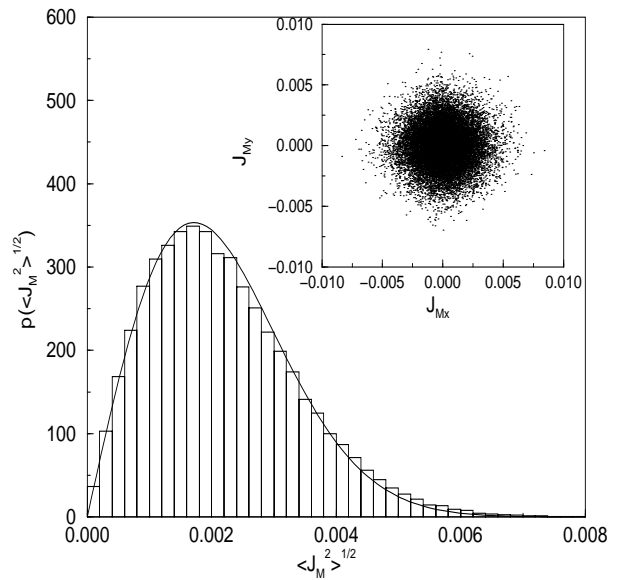,height=3.5in,width=3.5in,angle=0}}
\vspace{0.3cm}
\vskip -.5cm
\caption{Probability density of current distribution (columns) as compared to the Gaussian distribution (solid line). Inset: Scattering histogram of currents for $20,000$ independent samples with $M=26, h_0=1.0, \sigma_f=5.0, E=-3.0$ and $W=4.0$.}
\label{J2}
\end{figure}

The calculation is the same as described in Section~\ref{sec:wavefunction}, except we switched back to $h_0=1.0$ and $\sigma_f=5.0$, and applied atop a scalar potential randomly chosen within $[-\frac{1}{2}W,\frac{1}{2}W]$. The sample size is $M=26,36,50,70,100,140$, and $200$. The number of samples is typically $10^4$ or higher to reach a precision of $2-3\%$. In the inset of Figure~\ref{J2}, $20,000$ data points are shown, each dot representing the current $(J_x,J_y)$ of a configuration of magnetic field and scalar potential with $M=26$, $W=4.0$ and at Fermi energy $E=-1.0$. Much to our expectation, the statistical distribution of currents is symmetric around the origin $\vec J=0$ and appears to satisfy a Gaussian distribution. From the data in the inset, the probability density of finding a current at a given amplitude $|J_M|=\sqrt{J_{Mx}^2+J_{My}^2}$ is computed and shown in the main panel of Fig.~\ref{J2}. The solid line is the Gaussian distribution function, plotted for comparison:
\[p(x)=\frac{1}{\sigma^2}\,x\,e^{-\frac{x^2}{2\sigma^2}}\]
where $x\equiv |J_M|$ and $\sigma\equiv\sqrt{\langle J_M^2\rangle}$. With a very good agreement otained in Fig.~\ref{J2}, the rms current is thus a self-averaging quantity; that is, its average over $N$ independent samples converges as $\frac{1}{\sqrt N}$ for large $N$. Therefore, in our following computation, we shall actually take $\sqrt{\langle J_M^2\rangle}=\sqrt{\frac{\pi}{2}}\,\langle|J_M|\rangle$ for convenience.

\begin{figure}[htb]
\centerline{\epsfig{file=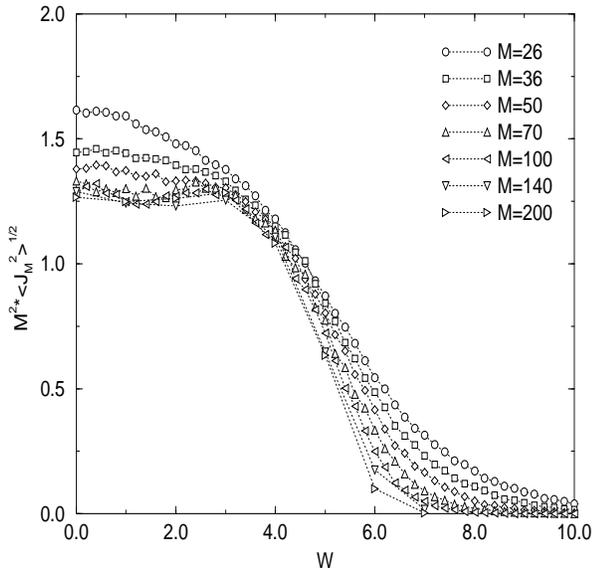,height=3.5in,width=3.5in,angle=0}}
\vspace{0.3cm}
\vskip -1cm
\caption{Root-mean-square of current.}
\label{J_W}
\end{figure}

Figure~\ref{J_W} presents similarly obtained results at $E=-1.0$ for various sample sizes and disorders. The initial impression of the picture is that $M^2\sqrt{\langle J_M^2\rangle}$ appears to be well-behaved. This justifies our use of $M^2\sqrt{\langle J_M^2\rangle}$ and, hence, our interpretation regarding it as a mearsure of local currents. Interestingly, its amplitude is also of order $t$ - the hopping element, exactly what expected of a current. We thus have a legal basis to regard $M^2\sqrt{\langle J_M^2\rangle}$ as the average local current, or local conduction, from now on. In the region of weak disorder, $W\alt 4.0$, it endures a slight size-dependence then tends to saturate at a finite value of order $t$ at large system size. This means that the bulk conduction on the links of the network survives the thermodynamic limit. On the other hand, in the region of strong disorder, it is strongly supressed as the system size is increased. The bulk local current is thus supposed to have undergone a transition from a finite value to zero as a function of disorder. The data in this computation, unfortunately, are not of good enough quality to locate a clear-cut transition point. Nonetheless, we see that the picture does indicate a critical disorder $W_c\approx 4.0$, consistent with our localization length study in Section~\ref{sec:twoparameter}.

Sheng and Weng, and Yang and Bhatt shortly afterward~\cite{ShengBhatt}, used the Chern number of a sample wrapped on a torus as a criterion of delocalization. By studying the scaling behavior of the number of current-carrying states as a function of sample size, these authors indeed identified a metal-insulator transition at a finite energy $E_c$, or equivalently a critical disorder $W_c$. Our results are compatible with their findings. However, there is a minor difference. While the Chern number is a global property of the wavefunction, we in fact dealt with its local attribute. Of course, one might argue that the local conduction is simply a reflection of the wavefunction as a whole, in that if the system is localized, the local current must be also small as a consequence. That is to say, our consideration of $M^2\sqrt{\langle J_M^2\rangle}$ delivers no new information as does the localization length. We, equiped by the justification of $M^2\sqrt{\langle J_M^2\rangle}$ as a local conduction, have looked at the problem from below. In a RMF system, the metal-insulator transition observed at large length scale should be regarded as the manisfestation of a transition of the conduction at microscopic distances.

Interestingly, the local current used in our work turns out to be analogous to the quantity $M^2\Delta E$ proposed by Edwards and Thouless~\cite{Edwards}, where $\Delta E$ is the energy shift as one switches from periodic to antiperiodic boundary condition. Their criterion of localization is to be adapted for the RMF case in what follows. Let us take an $M\times M$ square sample with some configuration of random magnetic field and scalar potential, and compute the current $J_M$ flowing through it at a given energy $E$. Suppose we now put many copies of the sample together to form an infinite periodic system with the original sample as a unit cell. Clearly, the current in the whole system is still $J_M$. We now allow these large unit cells, instead of being identical, to pick up different field configurations within the same population. Clearly, the average quantity $\sqrt{\langle J_M^2\rangle}$ would play the role of an ``effective'' hopping element, whereas the energy fluctuation among states around $E$ is typically $W/M^2$. The system is then equivalent to the old system with $t/W$ replaced by $M^2\sqrt{\langle J_M^2\rangle}/W$. If this parameter is smaller than the original one, it should obviously become even smaller if the argument is repeated one stage further, combining cells for bigger cells, and so on. Therefore, we arrive at a tentative criterion for localization:
\[\frac{t}{W}>\frac{M^2\sqrt{\langle J_M^2\rangle}}{W}\]
or
\begin{equation}
M^2\sqrt{\langle J_M^2\rangle}<t=1.
\label{eq:Edwards}
\end{equation}
Although the coefficient in the right-hand side of Eq.~(\ref{eq:Edwards}) should not be taken seriously, to our surprise, the numerical results in Fig.~\ref{J_W} show a remarkable agreement: at the putative critical disorder $W_c=4.0$, $M^2\sqrt{\langle J_M^2\rangle}$ is approximately $1.08\pm 0.02$.

It is noteworthy that the above argument is a coarse-graining one in essence. Literally, it was a precursor to later developments of the scaling theory of localization. By the same token, in the RMF problem, it elucidates the role of the new length scale -- $\sigma_f$, the correlation length of the random field, or roughly the average length of the links in its corresponding network. As one coarse-grains the system starting from a microscopic length, if the links contain few vortices, by the time one reaches $\sigma_f$, the vortices will have been renormalized away, leaving the system in a critical state. The system will then look self-similar at all length scales. On the other hand, if too many vortices are present, they will become dominant at large distances, driving the system into a insulating state.

\section{Discussion}
\label{sec:discussion}
We are now in a position to discuss other existing analytical approaches to the RMF problem. The first of these is the field theoretical $\sigma$-model which concludes that an electron gas subjected to a $\delta$-correlated RMF belongs to the unitary class of localization~\cite{Aronov}, thereby only sustaining an insulating phase. Although in two recent papers~\cite{Mirlin}, it was argued that the same results should be reached for a RMF with long-range correlation, we are confident that the visual pictures produced throughout our study have told us something otherwise. We believe that it is almost a general rule of thumb that the presence of a magnetic field, uniform or nonuniform, inevitably confers a profound influence upon the behavior of quantum particles. The electron wavefunction, as perfectly visible in Section~\ref{sec:wavefunction}, manifests a radical change in its structure as the magnetic field quenches its kinetic energy. In particular, the electron wavefunction acquires a fractal character accompanied by a spontaneous formation of vortex currents; both of which should be considered as {\it built-in} effects of the magnetic field. Moreover, the $\sigma$-model undoubtedly fails to account for the existence of the hidden degree of freedom pointed out and discussed throughout our paper. In another recent $\sigma$-model study~\cite{Taras-Semchuk}, the long-range effects of magnetic fields was considered and claimed to have yield a metal-insulator transition with a single node in the $\beta$-function. However, our results have discordantly established a whole critical phase, in which the $\beta$-function loses its putative meaning.

\begin{figure}[htb]
\centerline{\epsfig{file=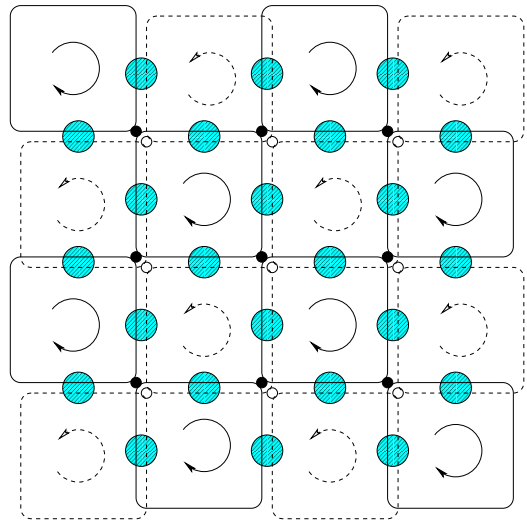,height=2.7in,width=2.7in,angle=0}}
\vspace{0.3cm}
\vskip 0.5cm
\caption{Schematic illustration of the two-channel network model. Clockwise and counterclockwise arrows indicate the current flows in boundaries of magnetic domains. Big shaded circles are the mixers where tunneling between two snake states (solid and dashed lines) takes place. Note that, Ref.18 does {\it not} allow snake states of different parities (even and odd) to mix at nodes (small solid and open circles) of the network.}
\label{2channel}
\end{figure}

As mentioned in a previous section, it is sensible to approximate the RMF problem by an effective network model. Actually, such an attempt has been made in Refs.~\cite{DKKLee,Kim}. However, in these studies, the possible role of drift states in localization was totally omitted. Rather, only the tunneling within pairs of snake states of even and odd parities (with respect to a reflection across their own zero-field line) was considered. We produce schematically in Fig.~\ref{2channel} such a two-channel network model in which the tunneling between a pair of channels takes place within the links of the network. The big squares represent magnetic domains of consecutive perpendicular directions. The arrows indicate the current flow along the boundaries between adjacent domains. While the mixing between snake states on the links is characterized as:
\[U=\left(\begin{array}{cc}
	e^{i\varphi_1} & 0\\
	0 & e^{i\varphi_2}
	\end{array} \right)
\left(\begin{array}{rr}
	\cos\phi & -\sin\phi\\
	\sin\phi & \cos\phi
	\end{array} \right)
\left(\begin{array}{cc}
	e^{i\varphi_3} & 0\\
	0 & e^{i\varphi_4}
	\end{array} \right), \]
in Refs.~\cite{DKKLee} the scattering at the nodes of the network, i.e., the saddle points of the magnetic field landscape, was parametrized as:
\begin{eqnarray}
	\left(\begin{array}{cccc}
	\cosh\theta_1 & 0 & \sinh\theta_1 & 0\\
	0 & \cosh\theta_2 & 0 & \sinh\theta_2\\
	\sinh\theta_1 & 0 & \cosh\theta_1 & 0\\
	0 & \sinh\theta_2 & 0 & \cosh\theta_2\\
\end{array} \right)
\label{node}
\end{eqnarray}

In virtue of the percolation at the saddle points, both $\theta_i$'s $(i=1,2)$ must be fixed equal to $\theta_c\equiv\ln(1+\sqrt 2)$, while their fluctuations around $\theta_c$ has been determined to be irrelevant~\cite{DHLee}. Using the above form of the scattering matrices, it was then asserted, by means of numerical simulations~\cite{DKKLee} and a spin-representation mapping~\cite{Kim}, that the model only supports an insulating phase. However, it is very important to note that, by means of (\ref{node}), the node scattering were incorrectly assumed to be {\it parity-preserving}. That is to say, as illustrated in Fig.~\ref{2channel}, snake states of even and odd parities were supposed to scatter independently at nodes. Without a parity-conservation rule at play, the scattering processes do {\it not} discriminate the two snake states. By that, the formulation in these studies~\cite{DKKLee,Kim} was too restrictive to draw a reliable outcome. One has to introduce an extra parameter in matrix~(\ref{node}) to describe the mixing between channels of opposite parities. In fact, the results obtained in Ref.~\cite{DKKLee} can be understood as follows. For two snake states with equal energy living on a link, the tunnneling between them lifts their degeneracy, effectively producing an additional potential fluctuations from link to link. The states would now find it difficult to tunnel through the nodes due to the restriction in the node scaterring, whose effect is expected to be nullified once more freedom in scattering is added.

In closing our discussion, there are two comments we wish to make:

(i) The two-channel network model can apply equally well to the problem of spin-unresolved QHE with the electron spins play the role of the two channels~\cite{DKKLee,Wang}. In this case, $\theta_i$'s are set equal to each other but can take an arbitratry value otherwise. It has been concluded~\cite{DKKLee,Wang} that the single critical point within a Landau band is split into two, the citical exponent at each point remaining unaffected $\nu=7/3$. At a further stage, if one now allows states of different spin to mix at the nodes of the network, it is interesting to ask whether the above picture would change, namely, whether the system would belong to another universality class.

(ii) One necessarily needs to introduce a new parameter characterizing the `blocking' processes on the links of the network. Note that the two-channel network model so far contained no adjustable parameter because $\theta_i$'s have already been fixed at criticality. We have actually carried out a computation for the standard Chalker-Coddington network model at criticality with an additional blocking factor on its links. Interestingly, the criticality is observed to be stable against blocking and the system undergoes a metal-insulator transtion of a new type. The results are to be reported elsewhere~\cite{own}.

\mbox{}

In conclusions, we have carried out the most comprehensive investigation of the localization properties of the RMF system. Our main findings, strongly supported by compelling evidences, are the existence of a finite region of extended states (critical phase) and a hidden degree of freedom. We devised a new two-parameter procedure to analyze our extensive simulation data. Not only does it recover the results of other well-understood situations, our method also elucidates localization problems from a new standpoint by testifying their very basis: the single-parameter scaling hypothesis. For the RMF model, the hypothesis, which has withstood the test of time, is found in our study to be invalidated by the presence of the hidden degree of freedom. In exploring a possible origin for this extra degree of freddom, we further established the dual role of the edge states that are formed along the magnetic field boundaries. On the one hand, extended states are carried by edge states by virtue of their percolation nature. On the other hand, specified by the hidden degree of freedom, the scattering among the edge states forms a new set of vortices, which, in turn, influences the electron transport in a fundamental way. Our proposed mechanism of localization in a RMF, that is, the tunneling between edge states (not the same mechanism as in other previous work, however), should deservedly be viewed as a counterpart to the coherent backscattering of weak localization.

Insofar as the possibility of conducting phases in disordered systems is concerned and given that quantum Hall states are the only known ones to sustain in two dimensions, the finding of a new conducting phase in our study is intriguing. In that vision, we hope that the RMF model would allow new theoretical perpectives to emerge, especially in light of paramount experimental evidences of an unexpected metallic groundstate in Si-MOSFET and other heterostructures~\cite{Kravchenko}. In particular, it would be a very interesting possibility if the RMF model would, arguably, turn out to be a non-trivial fixed point of disordered interacting systems under some circumstances~\cite{Chakravarty}.

\section{Acknowledgements}
I wish to thank Professor Sudip Chakravarty for his original inspiration and various critical comments in the work. I also thank Professor Steven Kivelson for his clarification on several important aspects of the network models, and Professor Hong-Wen Jiang for his explanations regarding numerous experimental results. The computations were performed at the computer lab of the Department of Physics and Astronomy, UCLA. This work was supported by a grant from the National Science Foundation, NSF-DMR-9971138.

\appendix

\section{Formation of vortices in a nonuniform magnetic field}
Consider a magnetic field perpendicular to the $x-y$ plane of the form $B(y)=B_0\,y$. Two types of classical orbits of electrons were shown in Fig.~\ref{Muller}, but the orbit size and the energy spectrum have to be determined by quantum mechanics. With the vector potential chosen as $\vec A=-\frac{1}{2}B_0\,y^2\,\hat x$, the $x$-component of momentum is a good quantum number: $p_x=\hbar k_x$. The wavefunction can thus be written in the form $\psi(x,y)=\chi(k_x,y)e^{ik_xx}$, where $\chi(k_x,y)$ is a solution to the equation
\begin{equation}
-\frac{\hbar^2}{2m}\,\chi''+[V_{eff}(k_x,y)-E]\,\chi=0.
\end{equation}
The effective potential
\begin{equation}
V_{eff}(k_x,y)=\frac{1}{2m}\left(\hbar k_x-\frac{eB_0}{2}\,y^2\right)^2
\end{equation}
is illustrated in Fig.~\ref{Veff} below.
\vskip .25cm

\begin{figure}[htb]
\centerline{\epsfig{file=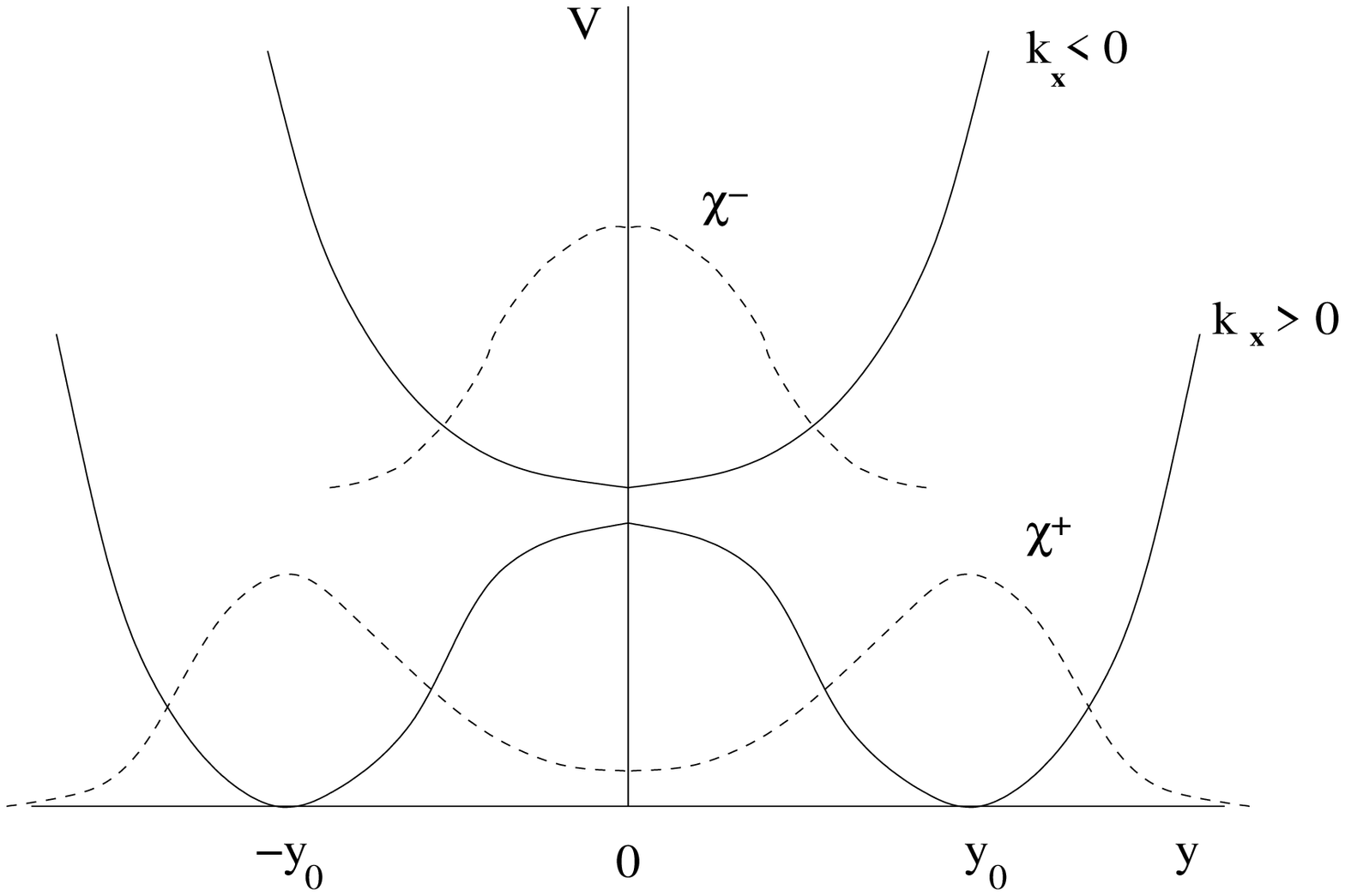,height=2.5in,width=3.5in,angle=0}}
\vspace{0.3cm}
\vskip -.3cm
\caption{}
\label{Veff}
\end{figure}

\begin{figure}[htb]
\centerline{\epsfig{file=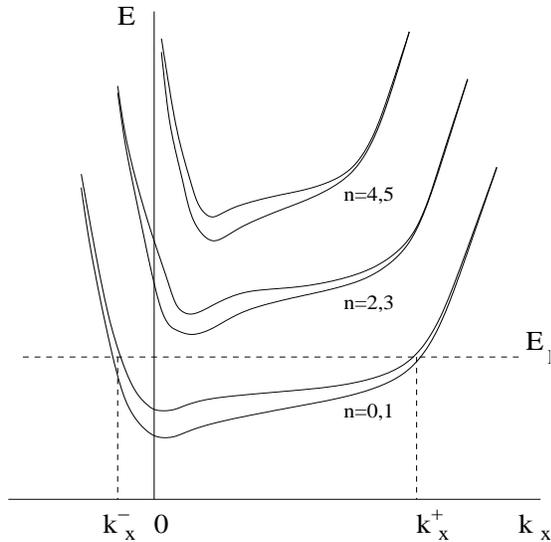,height=2.8in,width=2.9in,angle=0}}
\vspace{0.3cm}
\vskip 0cm
\caption{Energy spectrum of electron gas in a linearly varying magnetic field. Energy bands are formed pairwise with opossite parity.}
\label{Espectrum}
\end{figure}

For $k_x>0$, $V_{eff}$ has the form of a double well which supports symmetric and antisymmetric states about the $y$-axis. These states have peaks around the potential minima $\pm y_0=\pm (\frac{2\hbar k_x}{eB_0})^{1/2}$. Obviously, they are the {\it drift states} that reside on the two sides of the zero-field line and flow forwards in $\hat x$-direction. For $k_x\leq 0$, there are also symmetric amd antisymmetric states, but the split in their energy is bigger. These are the {\it snake states} that center on the zero-field line and, most of the time, flow backwards in $-\hat x$-direction. A schematic plot of the energy bands is shown in Fig.~\ref{Espectrum}. Beside $k_x$, the band index $n$ is also a good quantum number. It is indentified as the parity of $\chi$ with respect to a reflection about the $y$-axis. The energy bands appear to come with odd and even parities pairwise. Within such a pair of bands, except at an energy close to their minima, there are always two drift states and two snake states, each pair of states being very close in wavenumber. Therefore, it is sufficient to consider the scattering between one drift state and one snake state at wavenumber $k_x^+$ and $k_x^-$, respectively, which results in the following wavefunction:

\begin{equation}
\psi(x,y)=c^+\chi^+(y)\,e^{ik_x^+x}+c^-\chi^-(y)\,e^{ik_x^-x}
\label{eq:WF}
\end{equation}

\begin{figure}[htb]
\centerline{\epsfig{file=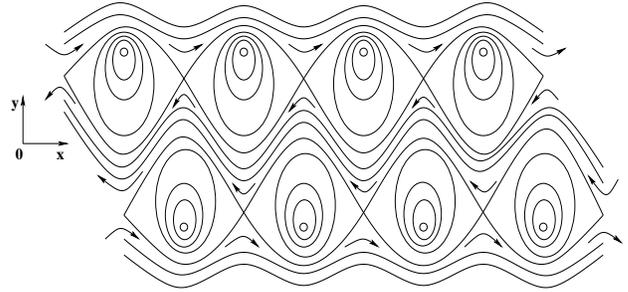,height=1.5in,width=3.2in,angle=0}}
\vspace{0.3cm}
\vskip 0.25cm
\caption{Schematic pattern of vortices.}
\label{vortices}
\end{figure}

Without loss of generality, we set $c^+=c^-=1$. Both $\chi^+(y)$ and $\chi^-(y)$ are {\it real} function of $y$. While $\chi^+(y)$ has two peaks at $\pm y_0$, $\chi^-(y)$ centers around $y=0$. Obviously, the wavefunction in Eq. (\ref{eq:WF}) has a series of isolated nodes at:
\[ \left\{
\begin{array}{lll}
x&=&\frac{(2n+1)\pi}{k_x^+-k_x^-} \\
y&=&y_1, \mbox{where $\chi^+(y_1)=\chi^-(y_1)$}
\end{array} \right. \]
and
\[ \left\{
\begin{array}{lll}
x&=&\frac{2n\,\pi}{k_x^+-k_x^-} \\
y&=&y_2, \mbox{where $\chi^+(y_2)=-\chi^-(y_2)$}
\end{array} \right. \]
The $y$-component of the current is:
\begin{equation}
j_y \propto (\frac{d\chi^+}{dy}\chi^--\frac{d\chi^-}{dy}\chi^+)\sin[(k_x^+-k_x^-)\,x]
\end{equation}
which vanishes at $x=n\,\pi/(k_x^+-k_x^-)$. A schematic picture of the wavefunction is shown in Fig.~\ref{vortices} where the direction of current flow is indicated by arrows. At the outer edges, there are forward flows corresponding to the drift states; while the snake states flowing backward in the center. There are also arrays of nodes and vortices in the interior of the wavefunction.
\vskip 0.5cm

It is noteworthy that the formation of vortices in a nonuniform magnetic field has not been mentioned before, to our understanding. The reason might have been that one only considered the clean case, i.e., the coupling between the snake and drift states was omitted. Several factors in real situations could induce the coupling, however. For example, they are finite length of the $B=0$ contour, its imperfect geometry, effect of impurities, to name a few.

\end{document}